\documentclass[a4paper,11pt]{article}

\usepackage{graphics,graphicx,epsfig,times,amsfonts,amssymb,amsmath,bbm,stmaryrd,latexsym}

\usepackage{hyperref}
\usepackage{color}
\hypersetup{colorlinks=true,linkcolor=blue,citecolor=magenta}

\newcommand{\cw}{}

\newcommand{\cb}{}

\newcommand{\cy}{}

\def\de{\delta}

\def\om{\omega}

\def\ve{\varepsilon}

\def\sfrac#1#2{{\textstyle\frac{#1}{#2}}}

\def\ph{\phantom{-}}
\def\ic{{\mathrm i}}
\def\im{{\mathrm i}}

\def\ep{{\mathrm e}}
\def\diff{{\mathrm d}}
\def\tr{{\mathrm tr}}

\def\>{\rangle}
\def\<{\langle}
\def\+{{\dagger}}
\def\we{{\wedge}}
\def\={\ =\ }

\newcommand{\Acal}{{\cal A}}
\newcommand{\Fcal}{{\cal F}}
\newcommand{\Ecal}{{\cal E}}
\newcommand{\Bcal}{{\cal B}}

\newcommand{\C}{\mathbb C}
\newcommand{\R}{\mathbb R}

\newcommand{\NN}{{\mathbb{N}}}

\newcommand{\beq}{\begin{equation}}
\newcommand{\eeq}{\end{equation}}
\newcommand{\bea}{\begin{eqnarray}}
\newcommand{\eea}{\end{eqnarray}}
\newcommand{\bal}{\begin{aligned}}
\newcommand{\eal}{\end{aligned}}

\newcommand{\with}{{\quad\textrm{with}\quad}}
\newcommand{\for}{{\quad\textrm{for}\quad}}
\renewcommand{\and}{{\quad\textrm{and}\quad}}
\newcommand{\und}{\qquad\textrm{and}\qquad}

\begin{document}

\ph
\vspace{1cm}

\begin{center}
{\huge\bf  Electromagnetic knots from de Sitter space}
\end{center}

\vspace{1cm}

\begin{center}
{\Large\bf Olaf Lechtenfeld}
\end{center}

\vspace{0.2cm}

\begin{center}

{\it
Institut f\"ur Theoretische Physik and
Riemann Center for Geometry and Physics\\
Leibniz Universit\"at Hannover,
Appelstrasse 2, 30167 Hannover, Germany}

\vspace{0.5cm}

\end{center}

\vspace{2cm}

\begin{abstract}
\noindent
We find all analytic SU(2) Yang--Mills solutions on de Sitter space 
by reducing the field equations to Newton's equation for a particle 
in a particular 3d potential and solving the latter in a special case.
In contrast, Maxwell's equations on de Sitter space can be solved
in generality, by separating them in hysperspherical coordinates.
Employing a well-known conformal map between (half of) de Sitter space 
and (the future half of) Minkowski space, the Maxwell solutions are 
mapped to a complete basis of rational electromagnetic knot configurations.
We discuss some of their properties and illustrate the construction method
with two nontrivial examples given by rational functions of increasing complexity.
The material is partly based on~\cite{lechtenfeld-zhilin,lechtenfeld-kumar}.
\end{abstract}

\vfill \noindent
Talk presented at the
RDP online workshop "Recent Advances in Mathematical Physics" - Regio2020,\\
06 December 2020, to appear in \href{https://pos.sissa.it/394/011/pdf}{PoS(Regio2020)011}.

\thispagestyle{empty}

\newpage
\setcounter{page}{1}

\section{Description of de Sitter space}

\noindent
Four-dimensional de Sitter space is a one-sheeted hyperboloid (of radius~$\ell$)
in $\R^{1,4}\ni\{Z_0,Z_1,\ldots,Z_4\}$ given by
\beq\cy
-Z_0^2 + Z_1^2 + Z_2^2 + Z_3^2 + Z_4^2 \=\ell^2\ .
\eeq
Constant $Z_0$ slices are 3-spheres of varying radius, 
yielding a parametrization of dS$_4\ni\{\tau,\omega_A\}$ as
\beq
\bal
\cy Z_0 \= -\ell\,\cot\tau \quad&\und\quad
Z_A \= \frac{\ell}{\sin\tau}\,\omega_A \for A = 1,\ldots,4 \\[4pt]
\with\qquad\cy \tau\in{\cal I}:=(0,\pi)
\quad&\und\quad \omega_A \omega_A = 1\ .
\eal
\eeq
The details of the embedding $\omega_A: (\chi,\theta,\phi)\ni S^3\hookrightarrow\R^4$ are irrelevant.
The Minkowski metric
\beq
\mathrm{d}s^2 \= -\mathrm{d}Z_0^2 + \mathrm{d}Z_1^2 + \mathrm{d}Z_2^2
+ \mathrm{d}Z_3^2 + \mathrm{d}Z_4^2
\eeq
induces on dS$_4$ the metric
\beq
\mathrm{d}s^2 \= \frac{\ell^2}{\sin^2\!\tau}
\bigl(-\mathrm{d}\tau^2 + \mathrm{d}\Omega_3^2\bigr)
\quad\with\quad\cw \diff\Omega_3^2 \for S^3\ ,
\eeq
showing that dS$_4$ is conformally equivalent to a finite cylinder ${\cal I}\times S^3$.

\bigskip

\section{Reduction of Yang--Mills to matrix equations}

\noindent
We wish to find solutions to the Yang--Mills (and Maxwell) equations on de Sitter space.
Due to their conformal invariance in four spacetime dimensions, we may also study the
problem on the finite Minkowskian cylinder~${\cal I}\times S^3$.

The gauge potential taking values in a Lie algebra~$\mathfrak{g}$ can always be chosen as
\beq\cy
\Acal \= X_a(\tau,\omega) \; e^a
\qquad\cw \textrm{on} \quad {\cal I}\times S^3
\eeq
where $X_a\in\mathfrak{g}$, and $\{e^a, a=1,2,3\}$ is a basis of left-invariant one-forms 
on $S^3\simeq\textrm{SU}(2)$, with
\beq\cy
\mathrm{d}e^a + \varepsilon^a_{\ bc}\,e^b\we e^c \=0 \quad\und\quad
e^a e^a \= \diff\Omega_3^2\ .
\eeq
There is no $\diff\tau$ component because we picked the temporal gauge $\Acal_\tau=0$.
In terms of the $S^3$ coordinates \ ($a,i,j,k=1,2,3$ and $B,C=1,2,3,4$) \ 
these one-forms can be constructed as
\beq\cy
e^a \= - \eta^a_{BC} \ \omega_B\,\mathrm{d}\omega_C
\qquad\textrm{\cw where}\quad\cy
\eta^i_{jk} = \varepsilon^i_{\ jk} \and \eta^i_{j4} = -\eta^i_{4j} = \delta^i_j\ .
\eeq
Dual to the $e^a$ are the left-invariant vector fields
\beq \label{Ralgebra}
R_a \= - \eta^a_{BC}\,\omega_B \frac{\partial}{\partial \omega_C}
\qquad\cw\Rightarrow\qquad\cy
[R_a,R_b] \= 2\,\varepsilon_{abc}\,R_c
\eeq
generating the right multiplication on SU(2), so that
an arbitrary function $\Phi$ on $S^3$ obeys
\beq
\mathrm{d}\Phi(\om) \= e^a\,R_a \Phi(\om)\ .
\eeq
The full SO(4) isometry group of~$S^3$ is generated by 
left-invariant~$R_a$ and right-invariant~$L_a$.

In this language, the gauge field two-form becomes 
($\dot{X}_a\equiv\sfrac{\diff}{\diff\tau}X_a$)
\beq
\begin{aligned}
\cy \Fcal &\cy \= \Fcal_{\tau a}\,e^\tau\we e^a + \sfrac12 \Fcal_{bc}\,e^b\we e^c \\
&\cy \= \dot{X}_a\,e^\tau\we e^a +
\sfrac12\bigl(R_{[b}X_{c]}-2\ve^a_{bc}X_a + [X_b, X_c]\bigr)\,e^b\we e^c\ ,
\end{aligned}
\eeq
where we define $R_{[b}X_{c]}=R_bX_c{-}R_cX_b$, 
and the Yang--Mills Lagrangian reads
\beq
\begin{aligned}
\cy {\cal L} &\cy\= \sfrac18\,\tr \Fcal_{\mu\nu}\Fcal^{\mu\nu}\=
-\sfrac14\,\tr\Fcal_{\tau a}\Fcal_{\tau a}\ +\ \sfrac18\,\tr\Fcal_{ab}\Fcal_{ab}\\
&\cy\= -\sfrac12\tr\bigl\{ \sfrac12\dot{X}_a\dot{X}_a-2X_a X_a
+\ve_{abc}X_a D_{[b} X_{c]} -\sfrac14(D_{[a} X_{b]})(D_{[a} X_{b]})\bigr\}
\end{aligned}
\eeq
with the short-hand \ $D_a:=R_a+X_a$. 
The Yang--Mills equations using (\ref{Ralgebra}) then take the form
\beq \cy \label{YMeom}
\begin{aligned}
\ddot X_a \=& -4\,X_a+ 2\,\ve_{abc}R_{[b} X_{c]} + R_b R_{[b} X_{a]} +  3\,\ve_{abc}\, [X_b, X_c] \\
&\ +2[X_b,R_b X_a] - [X_b,R_a X_b] - [X_a,R_b X_b] - \bigl [X_b, [X_a,X_b]\bigr] \\[4pt]
\=& -4\,X_a+ 2\,\ve_{abc}R_b X_c + R_b R_b X_a - R_a R_b X_b +  3\,\ve_{abc}\, [X_b, X_c] \\
&\ +2[X_b,R_b X_a] - [X_b,R_a X_b] - [X_a,R_b X_b] - \bigl [X_b, [X_a,X_b]\bigr] 
\end{aligned}
\eeq
with the Gauss law
\beq \label{YMgauss}
R_a\dot{X}_a + [X_a,\dot{X}_a] \= 0\ .
\eeq

\bigskip

\section{Yang--Mills configurations on de Sitter space}

\noindent
The simplest Yang--Mills solutions are most symmetric. To obtain them, let us impose SO(4) symmetry
by setting \ $X_a(\tau,\omega)=X_a(\tau)$.
The Yang--Mills equations then become ordinary matrix differential equations
\cite{lechtenfeld-ivanova, lechtenfeld-popov, lechtenfeld-popov2},
\beq \cy
\ddot{X}_a \= -4\,X_a+ 3\,\ve_{abc}\, [X_b, X_c] - \bigl [X_b, [X_a,X_b]\bigr]
\qquad\and\qquad [X_a,\dot{X}_a] \= 0\ .
\eeq
These three coupled ordinary differential equations for the three matrix functions~$X_a(\tau)$
are still too complicated. However, for the gauge group SU(2), these equations admit some analytic solutions.
So let us choose a spin-$j$ representation of $\mathfrak{g}=su(2)$ 
and introduce the three SU(2) generators~$T_a$,
\beq \cy
[T_b, T_c] \= 2\,\ve^a_{bc}T_a \und
\tr(T_aT_b)\=-4C(j)\,\de_{ab}\ \for
C(j)=\sfrac13\,j(j{+}1)(2j{+}1)\ .
\eeq
A simple ansatz for the matrices $X_a$ is
\beq \cy
X_1=\Psi_1 T_1\ \cw,\cy\quad
X_2=\Psi_2 T_2\ \cw,\cy\quad
X_3=\Psi_3 T_3
\quad\with \Psi_a=\Psi_a(\tau)\in\R\ .
\eeq
The resulting simplification of Yang--Mills Lagrangian density,
\beq
\cy {\cal L}
\= 4\,C(j)\,\bigl\{\sfrac14\dot{\Psi}_a\dot{\Psi}_a - (\Psi_1-\Psi_2\Psi_3)^2-
(\Psi_2-\Psi_3\Psi_1)^2 - (\Psi_3-\Psi_1\Psi_2)^2\bigr\}\ ,
\eeq
suggests an interpretation of $\{\Psi_a\}$ as the coordinates of a Newtonian particle in $\R^3$
moving in a potential
\beq \label{tetrapot}
\sfrac12 {\cal V}(\Psi)\ \=\
(\Psi_1-\Psi_2\Psi_3)^2\ \ +\ \ (\Psi_2-\Psi_3\Psi_1)^2\ \ +\ \ (\Psi_3-\Psi_1\Psi_2)^2\ .
\qquad\ph
\eeq
\begin{figure}[h!]
\centering
\includegraphics[scale=1.2]{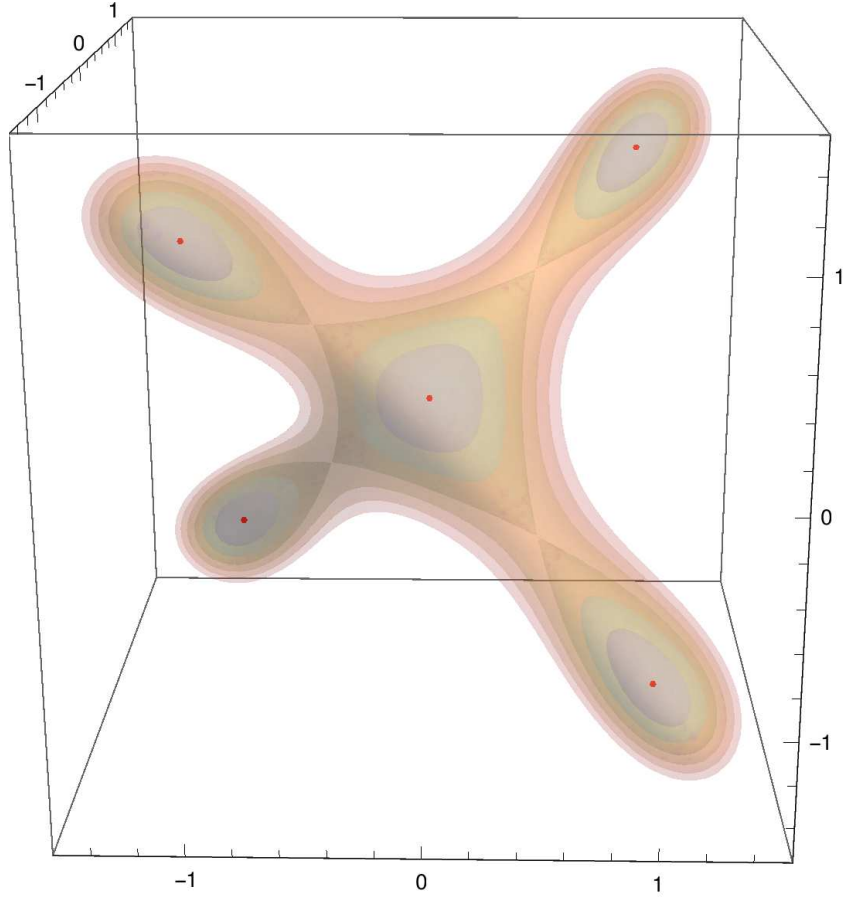}
\caption{Contours of the Newtonian potential in (\ref{tetrapot}).}
\label{fig1}
\end{figure}

The only analytic nonabelian solutions come from
\beq
\Psi_1=\Psi_2=\Psi_3=:\Psi 
\quad\with\quad \ddot{\Psi}=16\,\Psi\,(\Psi{-}1)(2\Psi{-}1) \ ,
\eeq
leading to elliptic functions $\Psi(\tau)$, except for the special cases 
$\Psi(\tau)=0 \  \textrm{or} \ 1$ (the vacuum),
$\Psi(\tau)=\sfrac12$ (the sphaleron),
and the bounce solution in the double-well potential.
The corresponding gauge potential takes the simple form
\beq
\Acal \= \Psi(\tau)\,g^{-1}\diff g \quad\for 
g:S^3 \buildrel{1:1}\over{\longrightarrow}\textrm{SU}(2)\ ,
\eeq
and the SU(2) color electric and magnetic fields are
\beq \cy
\Ecal_a \= \Fcal_{\tau a} \= \dot\Psi\,T_a \und
\Bcal_a \= \sfrac12\ve_{abc}\Fcal_{bc} \= 2\,\Psi\,(\Psi{-}1)\,T_a\ .
\eeq
Their total de Sitter energy and action is finite
and proportional to double-well energy.
These analytic Yang--Mills configurations are related to Minkowski-space solutions 
found in the seventies~\cite{dealfaro,luscher,schechter} 
(for a review from this period, see~\cite{actor}).
Their stability, however, has been analyzed only recently~\cite{KLP}.

\bigskip

\section{All Maxwell solutions on de Sitter space}

\noindent
The other analytic solutions to (\ref{YMeom}) and~(\ref{YMgauss}) are abelian,
i.e.~excite only a single direction in isospin space. 
In this case we can drop the matrix valuedness and treat the $X_a$ as real functions.
Dropping all commutator terms, the Yang--Mills equations~(\ref{YMeom}) turn into
the linear Mawell equations,
\beq \cy \label{Maxwelleom}
\ddot X_a \= (R^2-4)\,X_a+ 2\,\ve_{abc}R_b X_c
\eeq
where $R^2\equiv R_b R_b$ is the laplacian on~$S^3$, 
and we refined the temporal gauge to the Coulomb gauge
\beq
\Acal_\tau = 0 \und R_a X_a = 0\ ,
\eeq
which takes care of the Gauss law.

The coupled wave equations (\ref{Maxwelleom}) 
may be completely solved by separation of variables.
Seeking factorized complex basis solutions~\footnote{
$Z_a(\om)$ is not to be confused with the ambient-space coordinates~$Z_A$.}
\beq
X_a (\tau,\omega) \= Z_a(\om)\ \mathrm{e}^{\mathrm{i}\Omega\tau}\ ,
\eeq
one learns that the frequency $\Omega$ only depends on the SO(4) spin~$2j\in\NN_0$,
\beq
-R^2\,Z_a^j(\om) = 2j(2j{+}2)\,Z_a^j(\om) \qquad\Rightarrow\qquad
\bigl((\Omega^j)^2-4(j{+}1)^2\bigr)\bigl((\Omega^j)^2-4j^2\bigr) = 0\ ,
\eeq
where the second factor appears only for $j{\ge}1$.
The basis solutions~$Z_a^j$ to the linear system come in two types 
and carry two further labels $m$ and~$n$~\cite{lechtenfeld-zhilin}:
\begin{itemize}
\addtolength{\itemsep}{0pt}
\item type I : \quad
$j{\geq}0\ ,\quad m = -j,\ldots,+j\ ,\quad n = -j{-}1,\ldots,j{+}1\ ,\quad \cy\Omega^j=\pm2(j{+}1)$\\
\beq\cy \label{type1}
\begin{aligned}
Z_+^{j;m,n}  &\= \sqrt{(j{-}n)(j{-}n{+}1)/2} \ Y_{j;m,n+1} \\
Z_3^{j;m,n} \,&\= \sqrt{(j{-}n{+}1)(j{+}n{+}1)} \ Y_{j;m,n} \\
Z_-^{j;m,n}    &\= -\sqrt{(j{+}n)(j{+}n{+}1)/2} \ Y_{j;m,n-1}
\end{aligned}
\eeq
\item type II :\quad
$j{\geq}1\ ,\quad m = -j,\ldots,+j\ ,\quad n = -j{+}1,\ldots,j{-}1\ ,\quad \cy\Omega^j=\pm2j$\\
\beq\cy \label{type2}
\begin{aligned}
Z_+^{j;m,n}   &\= -\sqrt{(j{+}n)(j{+}n{+}1)/2} \ Y_{j;m,n+1} \\
Z_3^{j;m,n} \,&\= \sqrt{(j{+}n)(j{-}n)} \ Y_{j;m,n} \\
Z_-^{j;m,n}    &\= \sqrt{(j{-}n)(j{-}n{+}1)/2} \ Y_{j;m,n-1}
\end{aligned}
\eeq
\end{itemize}
where $Z_\pm=(Z_1\pm\im Z_2)/\sqrt{2}$, 
and the hyperspherical harmonics 
\beq\cy
Y_{j; m,n}(\omega) \qquad\with\quad m,n = -j,-j{+}1,\ldots,+j \und 2j=0,1,2,\ldots
\eeq
are characterized by~\footnote{
The label $m$ is the eigenvalue of $\sfrac{\im}{2}\,L_3$.}
\beq
-\sfrac14 R^2\,Y_{j; m,n}=j(j{+}1)\,Y_{j; m,n} \und
\sfrac{\im}{2}\,R_3\,Y_{j; m,n}=n\,Y_{j; m,n}\ .
\eeq

Hence, the general real Maxwell solution $\Acal=X_a(\tau,\omega) \; e^a$ is a linear combination with
\beq
X_a (\tau,\omega) \= \sum_{jmn} \Bigl\{
c^{\textrm{I}}_{j;m,n}\,Z_{a\ \textrm{I}}^{j;m,n}(\om)\ \mathrm{e}^{2(j+1)\,\mathrm{i}\tau}\ +\
c^{\textrm{II}}_{j;m,n}\,Z_{a\ \textrm{II}}^{j;m,n}(\om)\ \mathrm{e}^{2j\,\mathrm{i}\tau}\ +\ \textrm{c.c.}\Bigr\}\ .
\eeq
Each complex solution yields two real ones (real part and imaginary part). We count 
$2(2j{+}1)(2j{+}3)$ real type-I solutions and  $2(2j{+}1)(2j{-}1)$ real type-II solutions ($j{\ge}1$),
which add up to $4(2j{+}1)^2$ solutions for $j{>}0$ and 6 solutions for $j{=}0$, as it should.
Constant solutions ($\Omega=0$) are not allowed; 
the simplest ones are $j{=}0$ type~I or $j{=}1$ type~II. 
The most general $j{=}0$ configuration is
\beq
X_a^{(j=0)} \= \Bigl\{ 
c_{0;0,-1}\sfrac{1}{\sqrt{2}} \Bigl(\begin{smallmatrix} 1\\{-}\im\\0 \end{smallmatrix}\Bigr) +
c_{0;0,0} \Bigl(\begin{smallmatrix} 0\\0\\1 \end{smallmatrix}\Bigr) -
c_{0;0,+1}\sfrac{1}{\sqrt{2}} \Bigl(\begin{smallmatrix} 1\\\im\\0 \end{smallmatrix}\Bigr)
\Bigr\} \ \ep^{2\im\tau} \ +\ \textrm{c.c.}\ .
\eeq
The parity inversion, which interchanges left and right invariance, relates
spin $j$ \ type I \ solutions with spin $j{+}1$ \ type II \ solutions, swopping labels $m$ and~$n$.
Finally, electromagnetic duality is realized by shifting \ $|\Omega^j|\tau$ \ by \ $\pm\sfrac{\pi}{2}$, 
which produces from a solution~$\Acal$ a dual solution $\Acal_D$.
We shall now see that this basis of Maxwell solutions relates to so-called electromagnetic knots in Minkowski space.

\bigskip

\section{Conformal mapping to Minkowski space}

\noindent
The $Z_0{+}Z_4{<}0$ half of dS$_4$ is also conformally related to
future Minkowski space $\R^{1,3}_+\ni\{t,x,y,z\}$,
\beq
\bal &\cy
Z_0 = \frac{t^2{-}r^2{-}\ell^2}{2\,t}\ ,\quad
Z_1 = \ell\,\frac{x}{t}\ ,\quad
Z_2 = \ell\,\frac{y}{t}\ ,\quad
Z_3 = \ell\,\frac{z}{t}\ ,\quad
Z_4 = \frac{r^2{-}t^2{-}\ell^2}{2\,t} \\[4pt]
&\textrm{with}\qquad\cy
x,y,z\in\R \quad\und\quad r^2 = x^2 + y^2 + z^2 \qquad\textrm{\cw but}\qquad t\in\R_+\ ,
\eal
\eeq
since \ $t\in[0,\infty]$ \ corresponds to \ $Z_0\in[-\infty,\infty]$ \ but \ $Z_0{+}Z_4<0$.
In these Minkowski coordinates,
\beq
\mathrm{d}s^2 \=
\frac{\ell^2}{t^2}\,\bigl(-\mathrm{d}t^2 +\mathrm{d}x^2 +\mathrm{d}y^2 +\mathrm{d}z^2\bigr)\ .
\eeq
One may cover the entire $\R^{1,3}$ by gluing a second dS$_4$ copy and using the patch $Z_0{+}Z_4>0$.

We shall employ the direct relation between the cylinder and Minkowski coordinates,
\beq\cy \label{cyltomink}
\cot\tau = \frac{r^2 {-} t^2 {+} \ell^2}{2\,\ell\,t}\ ,\quad
\omega_1 = \gamma\,\frac{x}{\ell}\ ,\quad
\omega_2 = \gamma\,\frac{y}{\ell}\ ,\quad
\omega_3 = \gamma\,\frac{z}{\ell}\ ,\quad
\omega_4 = \gamma \frac{r^2 {-} t^2 {-} \ell^2}{2\,\ell^2}\ ,
\eeq
with the convenient abbreviation
\beq
\gamma \= \frac{2\,\ell^2}{\sqrt{4\,\ell^2 t^2 + (r^2-t^2+\ell^2)^2}}\ .
\eeq
Since $t=-\infty, 0, \infty$ \  corresponds to \ $\tau=-\pi,0,\pi$,
the cylinder gets doubled to \  $2{\cal I}\times S^3$,
and full Minkowski space is covered by the cylinder patch \ $\omega_4 \le\cos\tau$.
\begin{figure}[h!]
\centering
\includegraphics[scale=1.2]{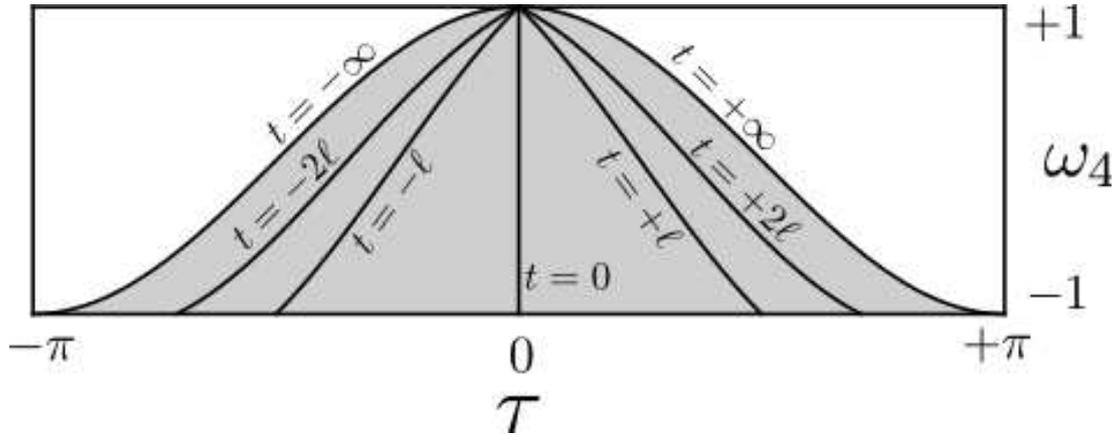}
\caption{An illustration of the map between a cylinder $2\mathcal{I}{\times}S^3$ and Minkowski space $\R^{1,3}$.
The Minkowski coordinates cover the shaded area. Its boundary is given by the curve $\omega_4 = \cos\tau$.
Each point is a two-sphere spanned by $\omega_{1,2,3}$, which is mapped to a sphere of constant $r$ and $t$.}
\label{fig2}
\end{figure}
The cylinder time $\tau$ is a regular smooth function of $(t,x,y,z)$, but more useful will be
\beq\cy
\exp (2\mathrm{i}\,\tau) \= \frac{\bigl[(\ell+\mathrm{i}t)^2+r^2\bigr]^2}{4\,\ell^2 t^2 + (r^2-t^2+\ell^2)^2}\ .
\eeq
A slightly lengthy computation yields the Minkowski-coordinate expressions for the one-forms
\cite{lechtenfeld-zhilin},
\beq\cy
\begin{aligned}
e^0 &\= e^0_\mu\,\diff x^\mu \= \frac{\gamma^2}{\ell^3}\Bigl(
\sfrac12(t^2 + r^2 + \ell^2)\,\mathrm{d}t - t\,x^k \mathrm{d}x^k \Bigr)\ ,\\[4pt]
e^a &\= e^a_\mu\,\diff x^\mu \= \frac{\gamma^2}{\ell^3}\Bigl(
t\,x^a \mathrm{d}t - \bigl(\sfrac12(t^2 - r^2 + \ell^2)\,\delta^a_k + 
x^a x^k + \ell\,\varepsilon^a_{\ jk}x^j \bigr)\,\mathrm{d}x^k \Bigr)\ ,
\end{aligned}
\eeq 
with the notation
\beq
(x^i)=(x,y,z) \qquad\and\qquad
(x^\mu)=(x^0,x^i)=(t,x,y,z)\ .
\eeq

Due to the conformal invariance of the Maxwell equations, 
our oscillatory solutions on the cylinder~$2\mathcal{I}{\times}S^3$ may be transferred
to a basis of Maxwell solutions on Minkowski space (with certain fall-off properties).
To accomplish this task, we only have to effect the coordinate change~\footnote{
The $S^2$ angular coordinates $(\theta,\phi)$ on both sides can be identified. 
The map $(\tau,\chi)\mapsto(t,r)$ realizes the Penrose diagram of Minkowski space
\cite{lechtenfeld-kumar}.}
\beq
\textrm{from} \qquad
(\tau,\omega)\sim(\tau,\chi,\theta,\phi) \qquad\textrm{to}\qquad
x\equiv(x^\mu)=(t,x,y,z)\sim(t,r,\theta,\phi)\ ,
\eeq
so that
\beq\cy
\Acal \= X_a(\tau(x),\omega(x))\,e^a(x) \= A_\mu(x)\,\diff x^\mu
\qquad\textrm{\cw yielding \quad $A_\mu(x)$} \with A_t\neq0\ ,
\eeq
\beq\cy
\mathrm{d}\Acal \= \dot{X}_a\,e^0\we e^a
-\varepsilon^a_{\ bc} X_a\,e^b \we e^c
\= \sfrac{1}{2} F_{\mu \nu}\,\diff x^\mu \we \diff x^\nu
\qquad\textrm{\cw yielding \quad $F_{\mu\nu}(x)$}\ .
\eeq
From this, we obtain electric and magnetic fields \
$\cy E_i=F_{i0} \and B_i=\sfrac{1}{2}\varepsilon_{ijk}F_{jk}$.
For the computation it is helpful to recognize that $\exp(2\im\tau)$ 
is a rational function of $t$ and~$r$. It follows that all physical quantities
(and the gauge potential) are rational functions of the Minkowski coordinates!

\bigskip

\section{All knot solutions on Minkowski space}

\noindent
As we shall see below, the simplest ($j{=}0$) solutions neatly reproduces the celebrated
Hopf-Ra\~nada electromagnetic knot~\cite{elknots,knotreview}.
From our construction, some general features of all knot solutions can be inferred.

Firstly, at spatial infinity (for $t$ fixed) all field strengths decay like $r^{-4}$,
but they fall off only as $(t{\pm}r)^{-1}$ along the light-cone.
Hence, the asymptotic energy flow is concentrated on past and future null infinity
and peaks on the light-cone of the spacetime origin. 
Secondly, the ``knot basis'' forms a complete set of finite-action configurations.
Of course, it does not contain plane waves.
Thirdly, the obvious conserved (in Minkowski time) quantities are helicity and energy,
\beq
h \= \sfrac12 \int \limits_{\R^3} \ \bigl( A\wedge F + A_D \wedge F_D \bigr)
\und
E \= \sfrac12 \int \limits_{\R^3} \!\diff^3\! x \ \bigl(\vec{E}^2 + \vec{B}^2\bigr)\ ,
\eeq
where the spatial integration is done at fixed~$t$.
Their common scale is determined by the amplitude of the solution, but their ratio
is fixed for the basis configurations.
Both quantities are best computed in the ``sphere frame'' at $t=\tau=0$,
\beq
F \= \mathcal{E}_a\,e^a \we e^0 + \sfrac{1}{2} \mathcal{B}_a\,\varepsilon^a_{\ bc}\,e^b\we e^c\ .
\eeq
Let us focus on type~I solutions of a fixed spin~$j$ and suppress these indices. 
For those one finds
\beq
\mathcal{E}_a \= -\mathrm{i}\Omega \sum_{mn}
c_{m,n}\,Z_a^{m,n}\,\mathrm{e}^{ \mathrm{i}\Omega\tau}+\,\textrm{c.c.} \und
\mathcal{B}_a \= -\Omega \sum_{mn}
c_{m,n}\,Z_a^{m,n}\,\mathrm{e}^{ \mathrm{i}\Omega\tau}+\,\textrm{c.c.}\ ,
\eeq
which yields
\beq
\sfrac12\bigl(\Ecal_a\Ecal_a+\Bcal_a\Bcal_a\bigr) \= 2 \Omega^2
\bigl|{\textstyle\sum}_{m,n}c_{m,n}\,Z_{a}^{m,n}(\om)\bigr|^2\ .
\eeq
The Minkowski energy at $t{=}0$ is easily pulled back to the cylinder frame
and evaluated by exploiting the orthogonality properties of the hyperspherical harmonics
\cite{lechtenfeld-kumar},
\beq
E \= \sfrac{1}{2\ell}\int \limits_{S^3} \!\!\diff^3\Omega_3 \  (1{-}\omega_4)\,
\bigl(\mathcal{E}_a\mathcal{E}_a + \mathcal{B}_a \mathcal{B}_a\bigr)
\=  \sfrac1\ell\,(2j{+}1)\,\Omega^3 \sum_{m,n} |c_{m,n}|^2\ .
\eeq
A similar computation produces an expression for the helicity.
It turns out that single-spin solutions (of both types)  have a universal energy-to-helicity ratio \
$E/h=|\Omega|/\ell$.

Fourthly, so-called null fields are easily characterized,
\beq
\vec{E}^2{-}\vec{B}^2=0=\vec{E}\cdot\vec{B} 
\qquad\cw\Leftrightarrow\cy\qquad (\vec{E}\pm\im\vec{B})^2=0
\qquad\cw\Leftrightarrow\cy\qquad \sum_{a} (\mathcal{E}_a\pm\im\mathcal{B}_a)^2=0\ .
\eeq
For fixed spin~$j$ and type~I we infer from above that
\beq
\mathcal{E}_a + \im\mathcal{B}_a \= -2\im\,\Omega\,
\sum_{mn} c_{m,n}\,Z_a^{m,n}(\omega)\,\mathrm{e}^{ \mathrm{i}\Omega\tau}
\qquad\textrm{(no c.c.!)}\ ,
\eeq
hence in such a sector we have~\cite{lechtenfeld-kumar}
\beq
F_{\mu\nu} \quad\textrm{null} \qquad\cw\Leftrightarrow\cy\qquad
\sum_{a} \Bigl(\sum_{mn} c_{m,n}\,Z_a^{m,n}(\omega) \Bigr)^2 = 0\ .
\eeq
Given the known form of the functions $Z_a^{m,n}(\om)$ we can expand 
this expression in hyperspherical harmonics and arrive at
$\sfrac16(4j{+}1)(4j{+}2)(4j{+}3)$ homogeneous quadratic equations 
for $(2j{+}1)(2j{+}3)$ complex parameters $c_{m,n}$.
This system is vastly overdetermined, but only $4j^2{+}6j{+}1$ equations
are independent, and thus we are still left with $2j{+}2$ free complex parameters
for the solution manifold, which is explicitly parametrized as follows
\cite{lechtenfeld-kumar},\footnote{
These are the generic solutions. There also exist special solutions 
with $c_{m,n}=0$ for $|n|\neq j{+}1$.}
\beq\cy
c_{m,n}(w,\vec{z}) \= {\scriptstyle\sqrt{\binom{2j+2}{j+1-n}}}\ 
w^{\frac{j+1-n}{2j+2}}\ \ep^{2\pi\im k_m\frac{j+1-n}{2j+2}}\ z_m
\quad\with w\in\C^* \and  \vec{z}\equiv\{z_m\}\in\C^{2j+1}
\eeq
and a choice of $2j{+}1$ integers $k_m\in\{0,1,\ldots,2j{+}1\}$ 
(one of which can be absorbed into $z_m$).
Given that the overall scale of the solutions is irrelevant,
the null fields form a complete-intersection projective variety 
of complex dimension $2j{+}1$ inside $\C P^{(2j+1)(2j+3)-1}$.
The simplest example occurs for spin $j{=}0$, where the single 
null-field relation $c_{_{0,0}}^2=2\,c_{_{0,-1}}c_{_{0,1}}$
defines a generic rank-3 quadric in $\C P^2$ or, alternatively,
a cone over $\C P^1$ lying in $\C^3$.

\bigskip

\section{Examples}

\noindent
We close with two concrete examples. 
First, the $j{=}0$ case represents SO(4)-symmetric Maxwell solutions in de Sitter space,
meaning $X_a(\tau,\omega)=X_a(\tau)$ thus $R_aX_b=0$ and trivializing (\ref{Maxwelleom}) to
\beq
\ddot X_a \= -4\,X_a \qquad\Rightarrow\qquad
X_a(\tau) \= \xi_a\,\cos\bigl(2(\tau{-}\tau_a)\bigr)\ ,
\eeq
which describes an ellipse in $\R^3$.\footnote{
Every solution $X_a(\tau)$ spontaneously breaks the SO(4) invariance
by the choice of integration constants $(\xi_a,\tau_a)$.}
We may always choose a frame where $\xi_3=0$ and $\tau_2=0$. 
The overall amplitude is irrelevant as all equations are linear,
and solutions can be superposed at will. Specializing to 
\beq
\xi_1=\xi_2=-\sfrac18 \and \tau_1=\sfrac\pi4  \qquad\Leftrightarrow\qquad
c_{0;0,-1}=c_{0;0,0}=0 \and c_{0;0,1}\in\im\R\ ,
\eeq
one has a null configuration with components
\beq\cy
X_1(\tau) = -\sfrac{1}{8} \sin 2\tau\ ,\quad
X_2(\tau) = -\sfrac{1}{8} \cos 2\tau\ ,\quad
X_3(\tau) = 0 \ .
\eeq
The result of a short computation yields
\beq\cy
\vec{E} + \mathrm{i} \vec{B} \= \frac{\ell^2}{\bigl((t-\mathrm{i}\ell)^2-r^2\bigr)^3}
\begin{pmatrix}
(x-\mathrm{i}y)^2-(t-\mathrm{i}\ell-z)^2 \\
\mathrm{i}(x-\mathrm{i}y)^2 + \mathrm{i}(t-\mathrm{i}\ell-z)^2 \\
-2\,(x-\mathrm{i}y)\,(t-\mathrm{i}\ell-z)
\end{pmatrix}\ .
\eeq
This is the announced Hopf--Ra\~nada electromagnetic knot~\cite{elknots,knotreview}. 
Our approach also yields its gauge potential.

Second, let us take the real part of the $(j;m,n)=(1;0,0)$ type~I basis solution.
Combining $\ep^{4\ic\tau}{+}\ep^{-4\im\tau}=2\cos4\tau$ 
and expressing $Y_{1;0,\star}$ from (\ref{type1}) in terms of~$\om_A$, we get
\beq\cy
X_\pm = -\sfrac{\sqrt{3}}{\pi}\,(\om_1{\pm}\im\om_2)(\om_3{\pm}\im\om_4)\,\cos 4\tau \und
X_3 = -\sfrac{\sqrt{6}}{\pi}\,(\om_1^2{+}\om_2^2{-}\om_3^2{-}\om_4^2)\,\cos 4\tau \ .
\eeq
This solution takes the explicit form (putting $\ell{=}1$)
{\small
\begin{equation*}
\bal
(E+&\im B)_x \=
\frac{-2\im}{ \left((t-\im)^2-x^2-y^2-z^2\right)^5} \ \times \\
\times\ &\Bigl\{ 2y +3 \im t y -x z +2 t^2 y + 2\im t x z-8x^2 y-8y^3+4yz^2 \\
&+\ 4\im t^3 y -6t^2 xz -8\im t x^2 y-8\im t y^3+4\im t yz^2 +10x^3 z+10xy^2 z -2xz^3 \\
&+\ 2(\im t x z+x^2 y+y^3+y z^2)(-t^2+x^2+y^2+z^2) +(\im t y- x z) (-t^2+x^2+y^2+z^2)^2\Bigr\}\ ,
\qquad\ \ {}
\eal
\end{equation*}
\beq \label{j1sol}
\bal
(E+&\im B)_y \=
\frac{2\im}{\left((t-\im)^2-x^2-y^2-z^2\right)^5} \ \times \\
\times\ &\Bigl\{ 2x +3 \im t x + y z+2 t^2 x-2\im t y z-8x^3-8xy^2 +4 xz^2 \\
&+\ 4 \im t^3 x +6 t^2y z -8\im t x^3-8\im txy^2 +4 \im t x z^2-10 x^2 yz-10y^3 z +2y z^3 \\
&+\ 2 (-\im t y z+x^3+x y^2+x z^2)(-t^2+x^2+y^2+z^2) +(\im t x+ y z)(-t^2+x^2+y^2+z^2)^2\Bigr\}\ ,
\qquad{}
\eal
\eeq
\begin{equation*}
\bal
(E+&\im B)_z \=
\frac{\im}{\left((t-\im)^2-x^2-y^2-z^2\right)^5} \ \times \\[4pt]
\times\  &\Bigl\{1+2\im t+t^2-11 x^2- 11 y^2 +3 z^2 +4 \im t^3-16\im t x^2-16\im t y^2+4\im t z^2 \\
&-\ t^4-2 t^2 x^2-2t^2 y^2-2t^2 z^2+11 x^4+22x^2y^2+10 x^2 z^2+11y^4-10 y^2 z^2+3z^4 \\
&+\ 2 \im t (t^2-3 x^2-3 y^2-z^2) (t^2-x^2-y^2-z^2) -(t^2+x^2+y^2-z^2)(-t^2+x^2+y^2+z^2)^2\Bigr\}\ .
\eal
\end{equation*}
}
Figures \ref{fig3} and \ref{fig4} below show $t{=}0$ energy density level surfaces and 
a particular magnetic field line. 
\begin{figure}[h!]
\centering
\includegraphics[scale=0.43]{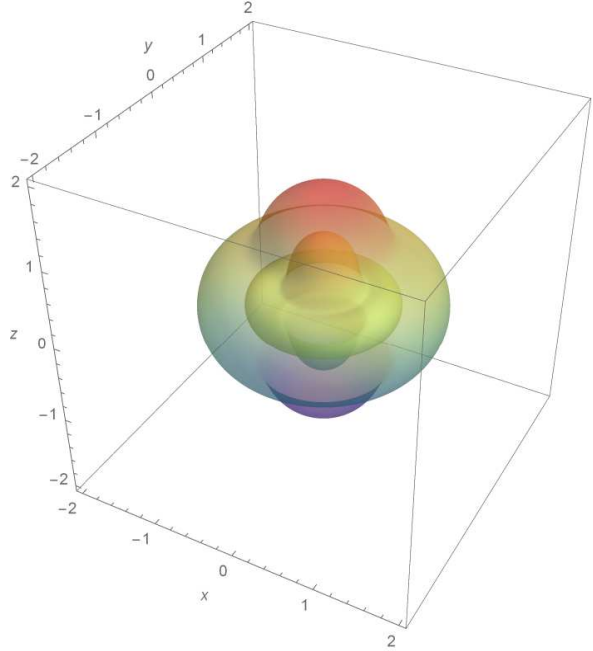}
\caption{Energy density level surfaces at $t{=}0$ for the $(1;0,0)$ solution above.}
\label{fig3}
\end{figure}
\begin{figure}[h!]
\centering
\includegraphics[scale=0.43]{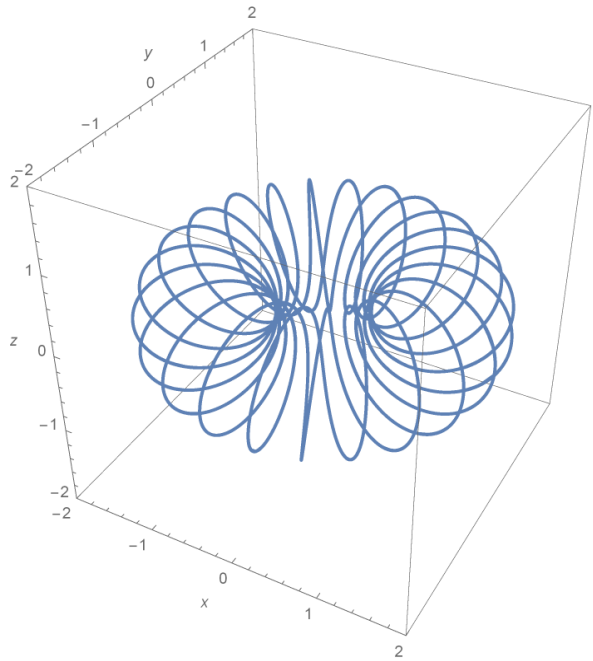}
\caption{A particular magnetic field line for the $(1;0,0)$ solution above.}
\label{fig4}
\end{figure}

\bigskip

\section{Summary and discussion}

\noindent
\begin{itemize}
\cb\item\cw
Rational electromagnetic fields with nontrivial topology have been investigated since 1989
\cb\item\cw
We introduced a new construction method based on two insights:
\\[-16pt]
\begin{itemize}
\cb\item\cw
the simplicity of solving Maxwell's equations on a temporal cylinder over a three-sphere
\cb\item\cw
the conformal equivalence of a cylinder patch $\{\tau,\omega\}$ to Minkowski space $\{x\}\equiv\{t,\vec{x}\}$
\end{itemize}
\cb\item\cw
The gauge potential is transferred via \ 
$\Acal=X_\nu(\tau,\omega)\,e^\nu=X_\nu(\tau(x),\omega(x))\,e^\nu_{\ \mu}(x)\,\diff x^\mu$
\cb\item\cw
Only finite-time $\tau\in(-\pi,+\pi)$ dynamics is required on the cylinder
\cb\item\cw
Our solutions have finite energy and action, by construction
\cb\item\cw
Energy and helicity are easily computed, null fields can be fully characterized
\cb\item\cw
A complete basis was constructed for sufficiently fast spatially and temporally decaying fields
\cb\item\cw
The non-Abelian extension couples different $j$ components of $X_a$ and will be harder to treat
\cb\item\cw
The method may be useful for numerics of Yang--Mills dynamics in Minkowski space
\end{itemize}

\vspace{1cm}

\noindent
{\bf\large Acknowledgments}\\[4pt]
The author is grateful to Gleb Zhilin and Kaushlendra Kumar 
for their contributions to \cite{lechtenfeld-zhilin,lechtenfeld-kumar}.



\end{document}